\documentclass[preprint,12pt]{elsarticle}

\usepackage{amssymb}

\usepackage{graphicx}   

\journal{Physica A}

\begin{document}

\begin{frontmatter}



\title{Quantum harmonic oscillator in option pricing}


\author{Liviu-Adrian Cotfas\fnref{lac}}
\address[lac]{Faculty of Economic Cybernetics, Statistics and Informatics, Academy of Economic Studies, 6 Piata Romana, 010374 Bucharest, Romania}
\ead{lcotfas@gmail.com}
\cortext[cor1]{Corresponding author}
\author{Nicolae Cotfas\corref{cor1}\fnref{nc}}
\address[nc]{University of Bucharest,  Physics Department, P.O. Box MG-11, 077125 Bucharest, Romania}
\ead{ncotfas@yahoo.com}

\begin{abstract}
The Black-Scholes model anticipates rather well the observed prices for options in the case of a strike price that is not too far from the current price of the underlying asset. Some useful extensions can be obtained by an adequate modification of the coefficients in the Black-Scholes equation. We investigate from a mathematical point of view an extension directly related to the quantum harmonic oscillator. In the considered case, the solution is the sum of a series involving the Hermite-Gauss functions. A finite-dimensional version is obtained by using a finite oscillator and the Harper functions. This simplified model keeps the essential characteristics of the continuous one and uses finite sums instead of series and integrals.

\end{abstract}


\end{frontmatter}


\section{Introduction}
\label{introd}
Black-Scholes (BS) equation, one of the most important equations in econophysics, is usually solved by
using its direct connection with the heat equation. The option price $V(S,t)$ at the moment $t$ 
is a function of the stock price $S$ at the considered moment of time. 
The function $V(S,t)$ can be expressed through a change of independent and dependent variable 
in terms of a function $u(x,t)$ satisfying the heat equation. The solution $V(S,t)$ of BS equation 
which coincides with the payoff function at the maturity time $T$ is directly obtained from a solution 
of heat equation satisfying a certain initial condition.

BS equation is very useful in finance, but it describes only an idealized case (constant volatility and risk-free interest rate). Some additional effects 
can be included by using an adequate modification of coefficients \cite{Baaquie}. In this paper, we consider only the 
case of a generalized Black-Scholes (GBS) equation obtained by adding a certain potential to the last coefficient.  In the case of a GBS equation a direct reduction to the heat equation seems to be impossible. The method of separation of variables can be used as an alternative tool. By looking for solutions having a particular form (separated variables), the GBS equation which is a differential equation with partial derivatives is reduced to a family of ordinary differential equations in one variable $x$ and depending on a parameter $\varepsilon $. More than that, for the considered GBS equations the corresponding equations in one variable are Schr\" odinger equations.

Two particular cases of GBS equation have been investigated by Jana and Roy in \cite{Jana}. Their purpose was to obtain a supersymmetric partner for each of them and to compute the pricing kernel in all the cases. We consider the case related to the quantum harmonic oscillator and we present:
\begin{itemize}
\item[-] a solution in terms of Hermite-Gauss functions,
\item[-] a class of supersymmetric partners depending on a continuous para-meter $\alpha $, and the corresponding solutions,
\item[-] a finite-dimensional approach based on a finite-difference operator, the finite Fourier transform and Harper functions.

\end{itemize}
\section{Black-Scholes equation}

We present, by following \cite{Ugur}, a short review concerning the use of the heat equation in order to offer the possibility to compare it with the approach based on the separation of variables. It is well-known that BS equation 
\begin{equation}\label{BS}
\frac {\partial V}{\partial t}+\frac{\sigma ^2}{2}S^2\frac{\partial ^2V}{\partial S^2}+rS\frac{\partial V}{\partial S}-rV=0
\end{equation}
for the European option price $V(S,t)$ is equivalent to the heat equation
\begin{equation}
\frac{\partial u}{\partial \tau }=\frac{\partial ^2u}{\partial x^2}
\end{equation}
whose solution for the initial condition $u(x,0)=u_0(x)$ is
\begin{equation}
u(x,\tau )=\frac{1}{\sqrt{4\pi \tau }} \int_{-\infty }^\infty {\rm e}^{-\frac{(x-\xi )^2}{4\tau }} u_0(\xi )\, d\xi .
\end{equation}
The solution of the Black-Scholes equation can be obtained in closed-form by using the change of independent and dependent variables 
\begin{equation}
\begin{array}{l}
S\!=\!K{\rm e}^x,\qquad
  t\!=\!T\!-\!\frac{2\tau }{ \sigma ^2},\qquad
V(S,t)\!=\!K\, {\rm e}^{-\gamma x-(\gamma +1)^2\tau }\, u(x,\tau )
\end{array}
\end{equation}
where
\begin{equation}
\begin{array}{l}
\gamma =\frac{r}{\sigma ^2}-\frac{1}{2}
\end{array}
\end{equation}
 $T$ is the maturity time, $S$ the stock price,  $K$ the strike price, $\sigma $ the volatility, and $r$ is the risk-free interest rate. In terms of the new variables, the payoff functions 
\begin{equation}
V_C(S,T)={\rm max}\{ S\!-\!K,0\},\qquad \quad V_P(S,T)={\rm max}\{ K\!-\!S,0\}
\end{equation}
become
\begin{equation}
u_C(x,0)\!=\!{\rm max}\{ {\rm e}^{(\gamma +1) x}\!-\!{\rm e}^{\gamma x},0\},\qquad \quad 
u_P(x,0)\!=\!{\rm max}\{ {\rm e}^{\gamma  x}\!-\!{\rm e}^{(\gamma +1) x},0\}
\end{equation}
and by denoting
\begin{equation}
\Phi (\zeta )=\frac{1}{\sqrt{2\pi }}\int_{-\infty }^\zeta {\rm e}^{-\eta /2}d\eta 
\end{equation}
the Black-Scholes formulas for the values of a European call and put option can be written in the closed form
\begin{equation}
\begin{array}{l}
V_C(S,t)=S\,  \Phi (d_1)-K\, {\rm e}^{-r(T-t)}\, \Phi (d_2)\\[2mm]
V_P(S,t)=K\, {\rm e}^{-r (T-t)}\, \Phi (d_1)-S\,  \Phi (d_2)
\end{array}
\end{equation}
with
\begin{equation}
\begin{array}{l}
d_1=\frac{{\rm log}(S/K)+\left(r+\frac{1}{2}\sigma ^2\right)(T-t)}{\sigma \sqrt{T-t}},\qquad 
d_2=\frac{{\rm log}(S/K)+\left(r -\frac{1}{2}\sigma ^2\right)(T-t)}{\sigma \sqrt{T-t}}.
\end{array}
\end{equation}
\section{A generalized version of Black-Scholes equation}

Let us  consider the more general version of Black-Scholes equation
\begin{equation}\label{gbs}
\frac {\partial V}{\partial t}+\frac{\sigma ^2}{2}S^2\frac{\partial ^2V}{\partial S^2}+rS\frac{\partial V}{\partial S}+(\sigma ^2\, U(\ln S)-r)V=0
\end{equation}
defined by using a  function $U:\mathbb{R}\longrightarrow \mathbb{R}$, called a potential \cite{Baaquie}. The function with separated variables
\begin{equation}\label{chvar}
V(S,t)={\rm e}^{\varepsilon t}\, S^{-\gamma }\, \phi (\ln S)
\end{equation}
where $\gamma $ is the constant used in the previous section, is a solution of equation (\ref{gbs}) 
if and only if $\phi (x)$ is a solution of the Schr\" odinger equation 
\begin{equation}
-\frac{1}{2}\, \frac{{\rm d}^2\phi }{{\rm d}x^2}+U(x)\, \phi =\lambda \, \phi 
\end{equation}
with
\begin{equation}\label{lambda}
\begin{array}{l}
\lambda =\frac{\varepsilon}{\sigma ^2}-\frac{1}{2}\left(\gamma \!+\!1\right)^2.
\end{array}
\end{equation}
The explicitly solvable cases 
\begin{equation}\label{twopot}
U(x)=0\qquad \mbox{and}\qquad 
U(x)=\left\{
\begin{array}{lll}
\infty & \mbox{for} & x\leq a\\
0 & \mbox{for} & a<x<b\\
\infty & \mbox{for} & x>b
\end{array}\right.
\end{equation}
have been analyzed in \cite{Jana} by using the factorization method. A supersymmetric partner has been obtained for each of them. Our purpose is to investigate the case $U(x)\!=\!\frac{x^2}{2}$ lying, in a certain sense, between the cases (\ref{twopot}). We present a solution in terms of Hermite-Gauss functions, a family of supersymmetric partners depending on a continuous parameter $\alpha $, and the corresponding solutions.

\section{A version related to the quantum harmonic oscillator}

The modified Black-Scholes equation
\begin{equation}\label{oscbs}
\frac {\partial V}{\partial t}+\frac{\sigma ^2}{2}S^2\frac{\partial ^2V}{\partial S^2}+rS\frac{\partial V}{\partial S}+\left(\frac{\sigma ^2}{2}(\ln S)^2-r\right)V=0
\end{equation}
is directly related to the Schr\" odinger equation of the quantum oscillator
\begin{equation}
-\frac{1}{2}\, \frac{{\rm d}^2\phi }{{\rm d}x^2}+\frac{x^2}{2}\phi =\lambda \, \phi .
\end{equation}
The function 
\begin{equation}
\Psi _n:\mathbb{R}\longrightarrow \mathbb{R},\qquad \Psi _n(x)=\frac{1}{\sqrt{n!\, 2^n\sqrt{\pi}}}\, H_n(x)\, {\rm e}^{-\frac{1}{2}x^2}
\end{equation}
where $H_n$ is the Hermite polynomial
\begin{equation}
H_n(x)=(-1)^n{\rm e}^{x^2}\, \frac{d^n}{dx^n}\left({\rm e}^{-x^2}\right)
\end{equation}
satisfies the equation
\begin{equation}
-\frac{1}{2}\, \frac{{\rm d}^2\Psi _n }{{\rm d}x^2}+\frac{x^2}{2}\Psi _n =\left(n\!+\!\frac{1}{2}\right) \Psi _n\qquad \mbox{for any}\quad  n\!\in \!\{ 0,1,2,...\}.
\end{equation}
The system of Hermite-Gaussian functions $\{\Psi _n\}_{ n\!\in \!\{ 0,1,2,...\}}$ is orthonormal
\begin{equation}
\int_{-\infty }^\infty \Psi _n(x)\, \Psi _k(x)\, dx=\delta _{nk}=\left\{ 
\begin{array}{lll}
1 & \mbox{if} & n=k\\
0 & \mbox{if} & n\neq k
\end{array}\right.
\end{equation}
and complete in the Hilbert space of  square integrable functions
\begin{equation}
L^2(\mathbb{R})=\left\{\, \psi :\mathbb{R}\longrightarrow \mathbb{C}\ \left|\ \ \int_{-\infty }^\infty |\psi (x)|^2dx<\infty \ \right. \right\}.
\end{equation}
In view of the result obtained in the previous section, the function 
\begin{equation}
V_n(S,t)={\rm e}^{\varepsilon _nt}\, S^{-\gamma }\, \Psi _n (\ln S)
\end{equation}
where
\begin{equation}
\begin{array}{l}
\varepsilon _n=n\sigma ^2+\frac{\sigma ^2}{2}+\frac{\sigma ^2}{2}\left(\gamma \!+\!1\right)^2
\end{array}
\end{equation}
is a solution of equation (\ref{oscbs}). If the coefficients $c_n$ are such that the function
\begin{equation}\label{solform}
V(S,t)=S^{-\gamma }\sum_{n=0}^\infty c_n\, {\rm e}^{\varepsilon _nt}\,  \Psi _n (\ln S)
\end{equation}
exists and can be derived term-by-term then it is also a solution of (\ref{oscbs}).
We choose $0<a<b $ such that the interval $(a,b)$ is large enough to contain all the possible values of the stock price $S$, and consider the square integrable  payoff functions
\begin{equation}
v_C(S)=\left\{ 
\begin{array}{lll}
0 &  \mbox{for} & S<K\\
S\!-\!K &  \mbox{for} & K\leq S\leq b\\
0 &  \mbox{for} & S>b\\
\end{array}\right.
\end{equation}
\begin{equation}
v_P(S)=\left\{ 
\begin{array}{lll}
0 &  \mbox{for} & S<a\\
K\!-\!S &  \mbox{for} & a\leq S\leq K\\
0 &  \mbox{for} & S>K\\
\end{array}\right.
\end{equation}
Since
\begin{equation}
\delta _{nk}=\int_{-\infty }^\infty \Psi _n(x)\, \Psi _k(x)\, dx=\int_0^\infty \frac{1}{S}\, \Psi _n(\ln S)\, \Psi _k(\ln S)\, dS\, ,
\end{equation}
from the relation
\begin{equation}
v_C(S)=S^{-\gamma }\sum_{n=0}^\infty c_n\, {\rm e}^{\varepsilon _nT}\,  \Psi _n (\ln S)
\end{equation}
we get
\begin{equation}
\begin{array}{rl}
c_n & \!\!\! ={\rm e}^{-\varepsilon _nT}\int_0^\infty v_C(s)\, s^{\gamma -1}\,  \Psi _n (\ln s)\, ds\\[3mm]
 & \!\!\! ={\rm e}^{-\varepsilon _nT}\int_K^b (s\!-\!K)\, s^{\gamma -1}\,  \Psi _n (\ln s)\, ds.
\end{array}
\end{equation}
The solution of the equation (\ref{oscbs}) is
\begin{equation}
V(S,t)\!=\!S^{-\gamma }\sum_{n=0}^\infty  {\rm e}^{\varepsilon _n(t-T)}\,  \Psi _n (\ln S)
\int_K^b (s\!-\!K)\, s^{\gamma -1}\,  \Psi _n (\ln s)\, ds.
\end{equation}
In the case of the put option the solution of (\ref{oscbs}) is
\begin{equation}
V(S,t)=S^{-\gamma }\sum_{n=0}^\infty  {\rm e}^{\varepsilon _n(t-T)}\,  \Psi _n (\ln S)
\int_a^K(K\!-\!s)\, s^{\gamma -1}\,  \Psi _n (\ln s)\, ds .
\end{equation}

If the real  parameter $\alpha $ is such that $|\alpha |\!>\!\frac{1}{2}\sqrt{\pi }$ then the potential \cite{Mielnik}
\begin{equation}
U_\alpha  (x)=\frac{x^2}{2}-\frac{dg_\alpha }{dx}(x)
\end{equation}
where,
\begin{equation}
g_\alpha :\mathbb{R}\longrightarrow \mathbb{R},\qquad g_\alpha (x)=\frac{{\rm e}^{-x^2}}{\alpha +\int_0^x{\rm e}^{-{u}^2}d{u}}\, ,
\end{equation}
is a supersymmetric partner of  $U(x)\!=\!\frac{x^2}{2}$. The functions
\[
\varphi _0,\ \ \varphi _1\!=\!A\Psi _0,\ \ \varphi _2\!=\!A\Psi _1,\ \ \varphi _3\!=\!A\Psi _2,\, ...
\]
where
\begin{equation}
\begin{array}{l}
\varphi _0(x)={\rm e}^{-\frac{x^2}{2}} \exp \left(\int_0^xg_\alpha (u)\, du\right)
\end{array}
\end{equation}
and $A$ is the first order differential operator
\begin{equation}
\begin{array}{l}
A=\frac{1}{\sqrt{2}}\left(-\frac{d}{dx}\!+\!x\!+\!g_\alpha (x)\right)\, ,
\end{array}
\end{equation}
belong to $L^2(\mathbb{R})$ and are orthogonal. 
The corresponding orthonormal system
\[
\Phi _0\!=\!\frac{\varphi _0}{||\varphi _0||},\ \ \Phi _1\!=\!\frac{\varphi _1}{||\varphi _1||},\ \ \Phi _2\!=\!\frac{\varphi _2}{||\varphi _2||},\ \ ...
\]
is complete in $L^2(\mathbb{R})$ and \cite{NLCotfas,Mielnik}
\begin{equation}
-\frac{1}{2}\, \frac{{\rm d}^2\Phi _n }{{\rm d}x^2}+U_\alpha (x)\, \Phi _n =\left(n\!+\!\frac{1}{2}\right) \Phi _n\qquad \mbox{for any}\quad  n\!\in \!\{ 0,1,2,...\}.
\end{equation}
The solution of the modified Black-Scholes equation
\begin{equation}
\frac {\partial V}{\partial t}+\frac{\sigma ^2}{2}S^2\frac{\partial ^2V}{\partial S^2}+rS\frac{\partial V}{\partial S}+(\sigma ^2\, U_\alpha (\ln S)-r)V=0
\end{equation}
satisfying the condition $V(S,T)=v_C(S)$ is
\begin{equation}
V(S,t)=S^{-\gamma }\sum_{n=0}^\infty  {\rm e}^{\varepsilon _n(t-T)}\,  \Phi _n (\ln S)
\int_K^b (s\!-\!K)\, s^{\gamma -1}\,  \Phi _n (\ln s)\, ds
\end{equation}
and the solution satisfying the condition $V(S,T)=v_P(S)$ is
\begin{equation}
V(S,t)=S^{-\gamma }\sum_{n=0}^\infty  {\rm e}^{\varepsilon _n(t-T)}\,  \Phi _n (\ln S)
\int_a^K(K\!-\!s)\, s^{\gamma -1}\,  \Phi _n (\ln s)\, ds .
\end{equation}

\section{An approach based on a finite quantum oscillator}

The Hamiltonian of the quantum oscillator can be written as
\begin{equation}
H=-\frac{1}{2}D^2+\frac{1}{2}x^2,\qquad \mbox{where}\quad D\!=\!\frac{d}{dx}
\end{equation}
The inverse of the Fourier transform
\begin{equation}
\psi \mapsto F[\psi ],\qquad F[\psi ](x )
=\frac{1}{\sqrt{2\pi }}\int_{-\infty }^\infty {\rm e}^{-{\rm i}x \xi }\, \psi (\xi )\,  dx.
\end{equation}
is the adjoint transformation
\begin{equation}
\psi \mapsto F^+[\psi ],\qquad F^+[\psi ](x )
=\frac{1}{\sqrt{2\pi }}\int_{-\infty }^\infty {\rm e}^{{\rm i}x \xi }\, \psi (\xi )\,  dx
\end{equation}
and
\begin{equation}
 F\left[D\psi \right](x)\!=\!{\rm i}x\, F[\psi ](x ),\qquad 
 F\left[D^2\psi \right](x)\!=\!-x^2\, F[\psi ](x ).
\end{equation}
If in the last formula we put $F^+[\psi ]$ instead of $\psi $ then we get 
\begin{equation}
FD^2F^+[\psi ](x)=-x^2\, \psi (x)
\end{equation}
and we have
\begin{equation}
H=-\frac{1}{2}(D^2+FD^2F^+).
\end{equation}
In order to obtain a finite counterpart of $H$,  we consider a positive odd integer $d\!=\!2\ell \!+\!1$ and the set \cite{NCotfas}
\[
\begin{array}{l}
\mathcal{R}_d\!=\!\left\{ -\ell \sqrt{\kappa },\, (-\ell \!+\!1)\sqrt{\kappa },\, \dots \, , 
\, (\ell \!-\!1)\sqrt{\kappa },\, \ell \sqrt{\kappa }\right\}\qquad \mbox{with}\quad \kappa \!=\!\frac{2\pi }{d}.
\end{array}
\]
The space $l^2(\mathcal{R}_d)$ of all the functions $\psi :\mathcal{R}_d\longrightarrow \mathbb{C}$  considered with the inner product
\begin{equation}
\begin{array}{l}
\langle \psi _1,\psi _2 \rangle =\sum\limits_{n=-\ell }^\ell \overline{\psi _1(n\sqrt{\kappa })}\,\psi _2 (n\sqrt{\kappa })
\end{array}
\end{equation}
is a Hilbert space isomorphic to the  $d$-dimensional Hilbert space $\mathbb{C}^d$.
Since
\begin{equation}
\lim_{d\rightarrow \infty }\sqrt{\kappa }=0\qquad \mbox{and} \qquad \lim_{d\rightarrow \infty }(\pm \ell )\sqrt{\kappa }=\pm \infty
\end{equation}
we can consider that, in a certain sense,
\begin{equation}
\mathcal{R}_d\stackrel{\scriptstyle{d\rightarrow \infty }}{-\!\!-\!\!\!\longrightarrow }\mathbb{R}\qquad \mbox{and} \qquad  l^2(\mathcal{R}_d)\stackrel{\scriptstyle{d\rightarrow \infty }}{-\!\!-\!\!\!\longrightarrow }L^2(\mathbb{R}).
\end{equation}
Each function $\psi :\mathcal{R}_d\longrightarrow \mathbb{C}$ can be regarded as the restriction to $\mathcal{R}_d$ of a periodic function $\psi :\mathbb{Z}\sqrt{\kappa  }\longrightarrow \mathbb{C}$ with period $d\sqrt{\kappa  }$.

The inverse of the finite Fourier transform $l^2(\mathcal{R}_d)\longrightarrow l^2(\mathcal{R}_d):\, \psi \mapsto \mathcal{F}[\psi ]$, where
\begin{equation}
\begin{array}{l}
\mathcal{F}[\psi ](n\sqrt{\kappa })=\frac{1}{\sqrt{d}}\sum\limits_{k=-\ell }^\ell {\rm e}^{-\frac{2\pi {\rm i}}{d}nk}\, \psi  (k\sqrt{\kappa }).
\end{array}
\end{equation}
is the adjoint transformation $l^2(\mathcal{R}_d)\!\longrightarrow 
\!l^2(\mathcal{R}_d):\, \psi \!\mapsto \!\mathcal{F}^+[\psi ]$, defined by
\begin{equation}
\begin{array}{l}
\mathcal{F}^+[\psi ](n\sqrt{\kappa })=\frac{1}{\sqrt{d}}\sum\limits_{k=-\ell }^\ell {\rm e}^{\frac{2\pi {\rm i}}{d}nk}\, \psi (k\sqrt{\kappa }).
\end{array}
\end{equation}
The finite-difference operator $\mathcal{D}^2$, where
\begin{equation}
\mathcal{D}^2\psi (n\sqrt{\kappa })=\frac{\psi ((n\!+\!1)\sqrt{\kappa })-2\psi  (n\sqrt{\kappa })+\psi  ((n\!-\!1)\sqrt{\kappa } )}{\kappa  }
\end{equation}
is an approximation of $D^2$, and we have
\begin{equation}
\begin{array}{l}
\mathcal{F}\mathcal{D}^2\mathcal{F}^+\psi (n\sqrt{\kappa  })=\frac{d}{\pi }\, \left(\cos \frac{2\pi n}{d}-1\right)\psi (n\sqrt{\kappa  }).
\end{array}
\end{equation}
The finite-difference Hamiltonian
\begin{equation}
\mathcal{H}_d=-\frac{1}{2}(\mathcal{D}^2+\mathcal{F} \mathcal{D}^2\mathcal{F}^+)
\end{equation}
with the matrix
\begin{equation}\label{matrix}
\begin{array}{l}
-\frac{d}{4\pi }\left(2 ( \cos \frac{2 \pi n}{d} \!-\! 2) \delta_{nm}  \!+\! 
 \delta_{n, m + 1} \!+\! \delta_{n, m - 1} \!+\! 
 \delta_{n, m - 2  \ell } \!+\! \delta_{n,m + 2  \ell }
\right)_{-\ell \leq n,m\leq \ell }
\end{array}
\end{equation}
that is,
\[
-\frac{d}{4\pi }\!\left( \!
\begin{array}{ccccc}
2\cos \frac{2 \pi (-\ell ) }{d} \!-\! 4 \!&\! 1 & 0 & \!\cdots \! & 1\\[2mm]
1 & 2\cos \frac{2 \pi (-\ell +1)}{d} \!-\! 4 & 1 & \!\cdots \! & 0\\[2mm]
0 & 1 \!&\! 2\cos \frac{2 \pi (-\ell +2)}{d} \!-\! 4 & \!\cdots \! & 0\\[1mm]
\vdots & \vdots & \vdots & \!\ddots \! & \vdots \\[1mm]
1 & 0 & 0 & \!\cdots \!  \!&\! 2\cos \frac{2 \pi \ell }{d} \!-\! 4 \\
\end{array}\! \right)
\]
is a finite counterpart of the Hamiltonian $H$ and, in a certain sense \cite{Ba},
\[
\mathcal{H}_d\stackrel{\scriptstyle{d\rightarrow \infty }}{-\!\!-\!\!\!\longrightarrow }H.
\] 
The operators $H$ and $\mathcal{H}_d$ are both Fourier invariant
\begin{equation}
FH=HF\qquad \mbox{and}\qquad  \mathcal{F}\mathcal{H}_d=\mathcal{H}_d \mathcal{F}.
\end{equation}
The eigenvalues of $\mathcal{H}_d$ are distinct, and the normalized eigenfunctions $ h_m$ of $\mathcal{H}_d$, considered in the increasing order of the 
number of sign alternations, can be regarded as a finite version of Hermite-Gaussian functions $\Psi _0,\Psi _1,...,\Psi _{d-1}$. For example, we have \cite{Ba}
\begin{equation}
F\Psi _m=(-{\rm i})^m\,\Psi _m\qquad \mbox{and}\qquad \mathcal{F} h_m=(-{\rm i})^m\, h_m .
\end{equation}
The functions $h_m$, called {\em Harper functions} \cite{Ba}, are eigenfunctions of $\mathcal{H}_d$ corresponding to certain eigenvalues $\lambda _n$, that is,
\begin{equation}
\mathcal{H}_d\, h_m=\lambda _m\, h_m,\qquad \mbox{for any}\quad m\!\in \!\{0,1,2,...,d\!-\!1\}.
\end{equation}
The eigenvalues $\lambda _n$ and the functions $h_n$ are available only numerically by diagonalizing the matrix (\ref{matrix}). Nevertheless, they play an important role in the theory of fractional Fourier transform, optics and signal processing \cite{OZK}.

In practice, the option  price  is a discrete variable, not a continuous one. 
It is an integer multiple of a certain minimal quantity, a sort of quantum of cash (usually, 1/100 or 1/1000 of the currency unit). In our simplified approach, we distinguish only a finite number of possible values, namely,
\begin{equation}
{\rm e}^{-\ell \sqrt{\kappa }},\ \ {\rm e}^{(-\ell +1) \sqrt{\kappa }},\ \ \dots \ \ {\rm e}^{(\ell -1) \sqrt{\kappa }},\ \ {\rm e}^{\ell \sqrt{\kappa }}.
\end{equation}
An acceptable description is obtained for $d\!=\!2\ell \!+\!1$ large enough. In the case of the strike price
$K={\rm e}^{k \sqrt{\kappa }}$ we consider the payoff functions
\begin{equation}
v_C({\rm e}^{n \sqrt{\kappa }})=\left\{ 
\begin{array}{llr}
0 &  \mbox{for} & -\ell \leq n<k\\
{\rm e}^{n \sqrt{\kappa }}\!-\!{\rm e}^{k \sqrt{\kappa }} &  \mbox{for} & k\leq n\leq \ell 
\end{array}\right.
\end{equation}
\begin{equation}
v_P({\rm e}^{n \sqrt{\kappa }})=\left\{ 
\begin{array}{llr}
{\rm e}^{k \sqrt{\kappa }}\!-\!{\rm e}^{n \sqrt{\kappa }} &  \mbox{for} & -\ell \leq n\leq k\\
0 &  \mbox{for} & k< n\leq \ell 
\end{array}\right.
\end{equation}
and assume that the solutions of (\ref{oscbs}) can be approximated by functions of the form (see (\ref{lambda}) and (\ref{solform}))
\begin{equation}
V({\rm e}^{m \sqrt{\kappa }},t)={\rm e}^{-m \gamma \sqrt{\kappa }}\sum_{n=0}^{d-1} c_n\, {\rm e}^{\varepsilon _nt}\,  h_n (m \sqrt{\kappa })
\end{equation}
with
\begin{equation}
\begin{array}{l}
\varepsilon _n=\sigma ^2\lambda _n+\frac{\sigma ^2}{2}\left(\gamma \!+\!1\right)^2.
\end{array}
\end{equation}
Since $\langle h_n,h_m\rangle =\delta _{nm}$,  from $V(S,T)=v_P(S)$ we get
\begin{equation}
c_n={\rm e}^{-\varepsilon _nT} \sum_{q=k }^\ell \left({\rm e}^{q \sqrt{\kappa }}\!-\!{\rm e}^{k \sqrt{\kappa }}\right) {\rm e}^{q \gamma \sqrt{\kappa }}h_n (q \sqrt{\kappa })
\end{equation}
and the corresponding solution of (\ref{oscbs}) can be approximated by
\[
V_C({\rm e}^{m \sqrt{\kappa }},t)\!=\!{\rm e}^{-m \gamma \sqrt{\kappa }}\sum_{n=0}^{d-1}  {\rm e}^{\varepsilon _n(t-T)}\,  h_n (m \sqrt{\kappa })\sum_{q=k }^\ell \left({\rm e}^{q \sqrt{\kappa }}\!-\!{\rm e}^{k \sqrt{\kappa }}\right) {\rm e}^{q \gamma \sqrt{\kappa }}h_n (q \sqrt{\kappa }).
\]
The solution of (\ref{oscbs}) satisfying $V(S,T)=v_P(S)$ can be approximated by
\[  
V_P({\rm e}^{m \sqrt{\kappa }},t)\!=\!{\rm e}^{-m \gamma \sqrt{\kappa }}\sum_{n=0}^{d-1}  {\rm e}^{\varepsilon _n(t-T)}\,  h_n (m \sqrt{\kappa })\sum_{q=-\ell  }^k \left({\rm e}^{k \sqrt{\kappa }}\!-\!{\rm e}^{q \sqrt{\kappa }}\right) {\rm e}^{q \gamma \sqrt{\kappa }}h_n (q \sqrt{\kappa })
\]
In Fig. 1 we present the values of $V_C({\rm e}^{m \sqrt{\kappa }},t)$ (left hand side) and $V_P({\rm e}^{m \sqrt{\kappa }},t)$ (right hand side) for $t\!=\!3$ (squares), $t\!=\!4$ (rhombus) and maturity time $T\!=\!5$ (bullets) in the case $d\!=\!21$, $\sigma \!=\!0.25$, $r\!=\!0.03$ and a strike price $K\!=\!{\rm e}^{8 \sqrt{\kappa }}$. In a neighbourhood of the strike price, our results agree with those obtained by using the standard BS equation.

\begin{figure}[t]
\centering
\includegraphics[scale=0.7]{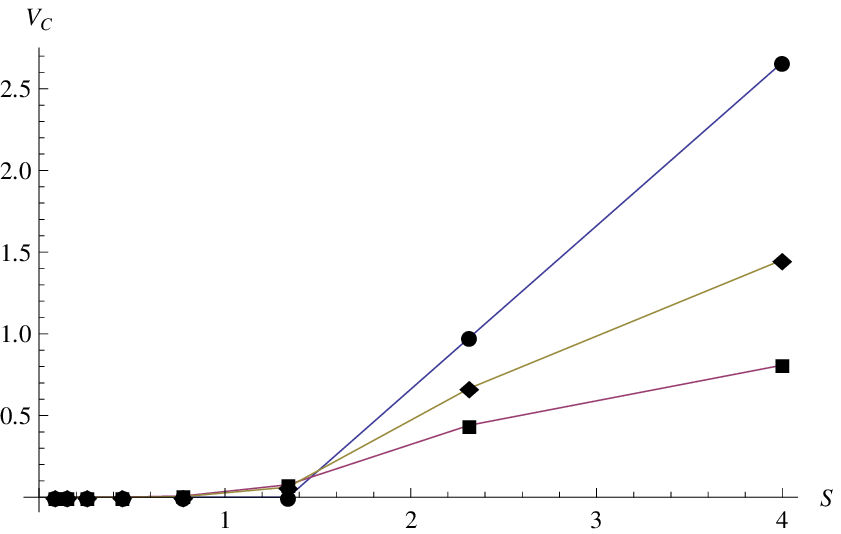}\qquad 
\includegraphics[scale=0.7]{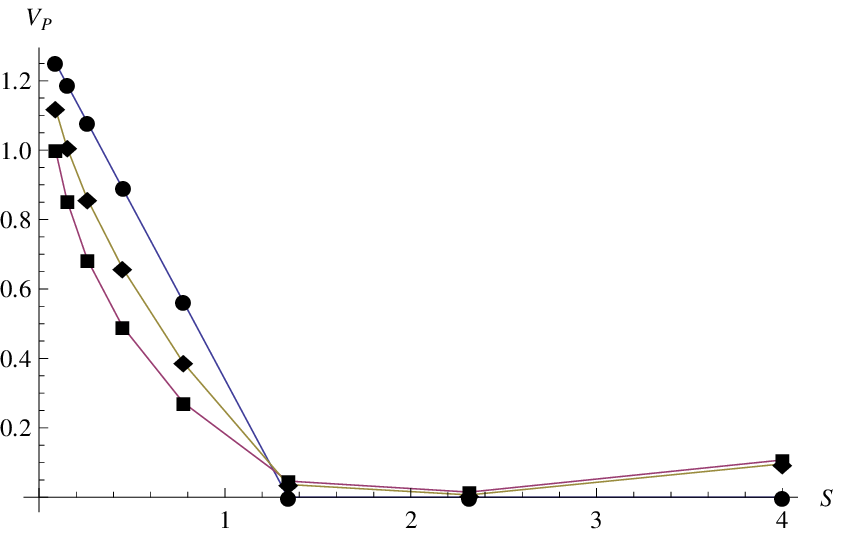}
\caption{\label{gaussians} Time evolution of the option price (see the text). }
\end{figure}

\section{Concluding remarks}
BS equation represents a `gold vein' for finance, and it is worth to dig around it because some other very interesting things may exist. We investigate from a mathematical point of view a modified version of the BS equation, without to know any possible financial interpretation. Our main contribution is a 
finite-dimensional approach to a GBS equation which keeps the essential characteristics of the  continuous case.

The mathematical modeling of price dynamics of the financial market is a very complex problem. 
We could never take into account all economic and non-economic conditions that have influences to the market \cite{De0,De}. Therefore, we usually consider some very simplified and idealized models, a kind of toy models which mimic certain features of a real stock market \cite{Bagarello2009,Bagarellobook}. We think that the finite-dimensional models \cite{Chen,LCotfas,NCotfas,Vourdas} may offer enough accuracy and are more accessible numerically . They use  linear operators with finite spectrum, finite sums instead of series and integrals, finite-difference operators instead of differential operators, etc.

The harmonic oscillator is among the most studied physical systems. Our results open a way to use in quantum finance the rich mathematical formalism developed around the quantum oscillator. We can use, for example, coherent states and the coherent state quantization \cite{Ga}, finite frames and the finite frame quantization \cite{NCotfas} in order to define mathematical objects with a financial meaning.








\end{document}